\documentclass[11pt,english,epsf]{article}
\usepackage{graphicx}
\usepackage{amsmath}
\usepackage{mathrsfs}
\usepackage{mathtools}
\usepackage{amssymb}
\usepackage{float}
\usepackage{url}
\usepackage{verbatim}
\usepackage{dsfont}
\usepackage{enumitem}
\usepackage{layout}

\newtheorem{theorem}{Theorem}

\newcommand {\bE} {\ensuremath{\mathbb{E}}}

\newcommand{\dfn}{\stackrel{\triangle}{=}}

\newcommand {\eqa} {\stackrel{\mbox{\tiny (a)}} {=}}

\newcommand {\eqc} {\stackrel{\mbox{\tiny (c)}} {=}}
\newcommand {\eqd} {\stackrel{\mbox{\tiny (d)}} {=}}

\newcommand {\lea} {\stackrel{\mbox{\tiny (a)}} {\le}}
\newcommand {\leb} {\stackrel{\mbox{\tiny (b)}} {\le}}
\newcommand {\lec} {\stackrel{\mbox{\tiny (c)}} {\le}}
\newcommand {\led} {\stackrel{\mbox{\tiny (d)}} {\le}}
\newcommand {\lee} {\stackrel{\mbox{\tiny (e)}} {\le}}
\newcommand {\lef} {\stackrel{\mbox{\tiny (f)}} {\le}}
\newcommand {\leg} {\stackrel{\mbox{\tiny (g)}} {\le}}

\newcommand {\gea} {\stackrel{\mbox{\tiny (a)}} {\ge}}
\newcommand {\geb} {\stackrel{\mbox{\tiny (b)}} {\ge}}

\newcommand {\gee} {\stackrel{\mbox{\tiny (e)}} {\ge}}
\newcommand {\gef} {\stackrel{\mbox{\tiny (f)}} {\ge}}

\newcommand{\hU}{{\hat{U}}}

\newcommand{\hS}{{\hat{S}}}
\newcommand{\hX}{{\hat{X}}}
\newcommand{\hY}{{\hat{Y}}}

\newcommand{\hW}{{\hat{W}}}
\newcommand{\hZ}{{\hat{Z}}}
\newcommand{\tb}{{\tilde{b}}}

\newcommand{\tu}{{\tilde{u}}}
\newcommand{\dw}{{\dot{w}}}
\newcommand{\dW}{{\dot{W}}}
\newcommand{\tv}{{\tilde{v}}}
\newcommand{\tw}{{\tilde{w}}}
\newcommand{\tx}{{\tilde{x}}}
\newcommand{\ty}{{\tilde{y}}}
\newcommand{\tz}{{\tilde{z}}}

\newcommand{\tX}{{\tilde{X}}}

\newcommand {\bu} {\mbox{\boldmath $u$}}

\newcommand {\bx} {\mbox{\boldmath $x$}}

\newcommand{\calA}{{\cal A}}

\newcommand{\calE}{{\cal E}}

\newcommand{\calS}{{\cal S}}

\newcommand{\calU}{{\cal U}}

\newcommand{\calW}{{\cal W}}
\newcommand{\calX}{{\cal X}}
\newcommand{\calY}{{\cal Y}}
\newcommand{\calZ}{{\cal Z}}
\allowdisplaybreaks

\begin{document}
\thispagestyle{empty}
\title{Encoding Individual Source Sequences for the Wiretap Channel
%\thanks{This research was supported by my wife and kids.}
}
\author{Neri Merhav
%\thanks{
%Currently on sabbatical leave at HP Laboratories,
%1501 Page Mill Road, MS 3U-4, Palo Alto CA 94304, USA.}
}
\date{}
\maketitle

\begin{center}
The Andrew \& Erna Viterbi Faculty of Electrical and Computer Engineering\\
Technion - Israel Institute of Technology \\
Technion City, Haifa 32000, ISRAEL \\
E--mail: {\tt merhav@ee.technion.ac.il}\\
\end{center}
\vspace{1.5\baselineskip}
\setlength{\baselineskip}{1.5\baselineskip}

\begin{abstract}
We consider the problem of encoding a deterministic source sequence (a.k.a.\ individual sequence) for the 
degraded wiretap channel
by means of an encoder and decoder that can both be implemented as finite--state machines. Our first main result is a necessary condition
for both reliable and secure transmission in terms of the given source sequence, the bandwidth expansion factor, the secrecy capacity,
the number of states of the encoder and the number of states of the decoder. Equivalently, this necessary condition can be presented as
a converse bound (i.e., a lower bound) on the smallest achievable bandwidth expansion factor. The bound is asymptotically achievable by
Lempel--Ziv compression followed by good channel coding for the wiretap channel. Given that the lower bound is saturated, we also derive a 
lower bound on the minimum necessary rate of purely random bits needed for local randomness at the encoder in order to meet the security
constraint. This bound too is achieved by the same achievability scheme. Finally, we extend the main results to the case where the
legitimate decoder has access to a side information sequence, which is another individual sequence that may be related to the source
sequence, and a noisy version of the side information sequence leaks to the wiretapper.
\end{abstract}

\section{Introduction}

In his seminal paper, Wyner \cite{Wyner75} has introduced the wiretap channel
as a model of secure communication
over a degraded broadcast channel, without using a secret key,
where the legitimate receiver has access to the output of the good channel
and the wiretapper receives the output of the bad channel. The main idea is that the excess noise
at the output of the wiretapper channel is utilized to secure the message intended to the legitimate receiver.
Wyner has fully characterized the best achievable trade--off
between reliable communication to the legitimate receiver and the equivocation rate at the
wiretapper, that was quantified in terms of
the conditional entropy of the source given the
output of the wiretapper channel. One of the most important concepts, introduced by Wyner, was
the {\em secrecy capacity}, that is, the supremum of all coding rates that
allow both reliable decoding at the
legitimate receiver and full secrecy, where the equivocation 
rate saturates at the (unconditional) entropy rate of the source,
or equivalently, the normalized mutual information between the source and the wiretap channel output is vanishingly
small for large block-length.
The idea behind the construction of a good code for the wiretap channel
is basically the same as the idea of binning: One
designs a big code, that can be reliably decoded at the legitimate receiver,
which is subdivided into smaller codes that are fed by purely random bits that are unrelated to the
secret message. Each such sub-code 
can be reliably decoded individually by the wiretapper to its full capacity, 
thus leaving no further decoding capability for
the remaining bits, which all belong to the real secret message. 

During the nearly five decades that have passed since \cite{Wyner75} was published, the wiretap channel model
has been extended and further developed in many aspects. We mention here just a few.
Three years after Wyner, 
Csisz\'ar and K\"orner \cite{CK78} have extended the wiretap channel to a general broadcast channel that is not
necessarily degraded, allowing also a common message intended to both receivers. In the same year, 
Leung--Yan--Cheong
and Hellman \cite{LYCM78}, studied the Gaussian wiretap channel,
and proved, among other things, that its secrecy capacity
is equal to the difference between the capacity of the legitimate channel and that of the wiretap channel.
In \cite{OW85}, Ozarow and Wyner
considered a somewhat different model, known as the type II wiretap channel, 
where the channel to the legitimate receiver
is clean (noiseless), and the wiretapper can access a subset of the coded bits.
In \cite{Yamamoto89}, Yamamoto extended the wiretap channel
to include two parallel broadcast channels,
that connect one encoder and one legitimate decoder, and
both channels are being wiretapped by
wiretappers that do not cooperate with each other.
A few years later, the same author \cite{Yamamoto97}
has further developed the scope of \cite{Wyner75} in two ways: First, by
allowing a private secret key to be shared between the encoder
and the legitimate receiver, and secondly, by allowing a given
distortion in the reproducing the source at
the legitimate receiver. The main coding theorem of \cite{Yamamoto97}
suggests a three-fold separation principle, which asserts that
no asymptotic optimality is lost if the encoder first, applies a
good lossy source code, then encrypts the compressed
bits, and finally, applies a good channel code for the wiretap channel.
In \cite{me08}, this model in turn was generalized to allow source side information at the decoder and at the
wiretapper in a degraded structure with application to systematic coding for the wiretap channel.
The Gaussian wiretap channel model of \cite{LYCM78} was also extended in two ways: The first is
the Gaussian multiple access wiretap channel of \cite{TY08}, and the second 
is Gaussian interference wiretap channel
of \cite{Mitrpant03}, \cite{MHVL04}, where the encoder has access to the interference signal as side information.
Wiretap channels with feedback were considered in \cite{AFJK09}, where it was shown that feedback is best used for
the purpose of sharing a secret key as in \cite{Yamamoto97} and \cite{me08}.
More recent research efforts were dedicated to strengthening the secrecy metric from weak secrecy to
strong secrecy, where the mutual information between the source and the wiretap channel output vanishes even
without normalization by the block-length, as well as to semantic security, which is similar but 
refers even to the worst--case message source
distribution, see, e.g., \cite[Section 3.3]{BB11}, \cite{BTV12}, \cite{GCP16}.

In this work, we look at Wyner's wiretap channel model from a different perspective. Following the
individual--sequence approach pioneered by Ziv in \cite{Ziv78}, \cite{Ziv80} and \cite{Ziv84}, and continued in later works, such as
\cite{me13} and \cite{me14}, we consider the problem of encoding a deterministic source 
sequence (a.k.a.\ an individual sequence) for the degraded wiretap channel using
finite--state encoders and finite--state decoders. One of the non-trivial issues associated with individual sequences, in the context
of the wiretap channel, is how to define the security metric, as there is no probability distribution assigned to the source, and therefore,
the equivocation, or the mutual information between the source and the wiretap channel output cannot be well defined.
In \cite{me13}, a similar dilemma was encountered in the context of private-key encryption of individual sequences, and in the
converse theorem therein, it was assumed that the system is perfectly secure in the sense that the probability distribution of the
cryptogram does not depend on the source sequence. 
In principle, it is possible to apply the same approach here, where the word `cryptogram' is replaced by the `wiretap channel output'.
But in order to handle residual dependencies, which will always exist, it would be
better to use a security metric that quantifies those small dependencies. To this end, it makes sense to adopt the above mentioned
maximum mutual information security metric (or, equivalently the semantic security metric), where the maximum is over all input assignments.
After this maximization, this quantity depends only on the `channel' between the source and the wiretap channel output.

Our first main result is a necessary condition (i.e., a converse to a coding theorem)
for both reliable and secure transmission, which depends on: (i) 
the given individual source sequence, (ii) the bandwidth expansion factor, (iii) the secrecy capacity,
(iv) the number of states of the encoder, (v) the number of states of the decoder, (vi) the allowed bit error probability at
the legitimate decoder and (vii) the allowed maximum--mutual--information secrecy. Equivalently, this necessary condition can be presented as
a converse bound (i.e., a lower bound) to the smallest achievable bandwidth expansion factor. The bound is asymptotically achievable by
Lempel--Ziv (LZ) compression followed by a good channel coding scheme for the wiretap channel. Given that this lower bound is saturated, we then derive also a
lower bound on the minimum necessary rate of purely random bits needed for adequate local randomness at the encoder, in order to meet the security
constraint. This bound too is achieved by the same achievability scheme, a fact which may be of independent interest
regardless of individual sequences and finite--state encoders and decoder (i.e., also for ordinary block codes in the traditional probabilistic setting). 
Finally, we extend the main results to the case where the
legitimate decoder has access to a side information sequence, which is another individual sequence that may be related to the source
sequence, and where a noisy version of the side information sequence leaks to the wiretapper. It turns out that in this case, the best strategy is the same as
if one assumes that the wiretapper sees the clean side information sequence. While this may not 
be surprising as far as sufficiency is concerned (i.e., as an achievability
result), it is less obvious in the context of necessity (i.e., a converse theorem). 

The remaining part of this article is organized as follows.
In Section \ref{ndps}, we establish the notation, provide some definitions and formalize the problem setting.
In Section \ref{results}, we provide the main results of this article and discuss them in detail.
In Section \ref{si}, the extension that incorporates side information is presented.
Finally, in Section \ref{proofs}, the proofs of the main theorems are given.

\section{Notation, Definitions, and Problem Setting}
\label{ndps}

\subsection{Notation}

Throughout this paper, random variables will be denoted by capital
letters, specific values they may take will be denoted by the
corresponding lower case letters, and their alphabets
will be denoted by calligraphic letters. Random
vectors, their realizations, and their alphabets will be denoted,
respectively, by capital letters, the corresponding lower case letters and caligraphic letters,
all superscriped by their dimensions.
For example, the random vector $X^n=(X_1,\ldots,X_n)$, ($n$ --
positive integer) may take a specific vector value $x^n=(x_1,\ldots,x_n)$
in $\calX^n$, the $n$--th order Cartesian power of $\calX$, which is
the alphabet of each component of this vector. Infinite sequences will be denoted using the bold 
face font, e.g., $\bx=(x_1,x_2,\ldots)$. Segments of vectors will be denoted by subscripts and superscripts that
correspond to the start and the end locations, for example, $x_i^j$, for $i < j$ integers, will denote $(x_i,x_{i+1},\ldots,x_j)$.
When $i=1$ the subscript will be omitted.

Sources and channels will be denoted by the letter $P$ or $Q$,
subscripted by the names of the relevant random variables/vectors and their
conditionings, if applicable, following the standard notation conventions,
e.g., $Q_X$, $P_{Y|X}$, and so on, or by abbreviated names that describe their functionality. When there is no room for ambiguity, these
subscripts will be omitted.
The probability of an event $\calE$ will be denoted by $\mbox{Pr}\{\calE\}$,
and the expectation
operator with respect to (w.r.t.) a probability distribution $P$ will be
denoted by
$\bE_P\{\cdot\}$. Again, the subscript will be omitted if the underlying
probability distribution is clear from the context or explicitly explained in the following text.
The indicator function of an event $\calE$ will be denoted by $1\{\calE\}$, that is, $1\{\calE\}=1$ if $\calE$ occurs,
otherwise, $1\{\calE\}=0$.

Throughout considerably large parts of the paper, the analysis will be carried out w.r.t.\ 
joint distributions that involve several random variables. Some of this random variables will be induced from empirical distributions
of deterministic sequences while others will be ordinary random variables. Random variables from the former kind will be denoted with `hats'.
As a simple example, consider a deterministic sequence, $x^n$,
that is fed as an input to a memoryless channel defined by a single--letter transition matrix, $\{P_{Y|X},~x\in\calX,~y\in\calY\}$, and let $y^n$
denote a realization of the corresponding channel output. Let $P_{\hX\hY}(x,y)=\frac{1}{n}\sum_{i=1}^n1\{x_i=x,~y_i=y\}$ denote the joint empirical 
distribution induced from $(x^n,y^n)$. In addition to $P_{\hX\hY}(x,y)$, we also define $P_{\hX Y}(x,y)=\bE\{P_{\hX\hY}(x,y)\}$, where now $Y$ is an
ordinary random variable. Clearly, the relation between the two distributions is given by $P_{\hX Y}(x,y)=P_{\hX}(x)\cdot P_{Y|X}(y|x)$, where
$P_{\hX}(x)=\sum_yP_{\hX\hY}(x,y)$ is the empirical marginal of $\hX$. Such mixed joint distributions will underlie certain information-theoretic quantities,
for example, $I(\hX;Y)$ and $H(Y|\hX)$ will denote, respectively, the mutual information between $\hX$ and $Y$ and the conditional entropy of $Y$ given $\hX$,
both induced from $P_{\hX Y}$. The same notation rules will be applicable in more involved situations too.

\subsection{Definitions and Problem Setting}

Let $\bu=(u_1,u_2,\ldots)$ be a deterministic source sequence 
(a.k.a.\ individual sequence), whose symbols take values in a finite alphabet,
$\calU$, of size $\alpha$. This source sequence is divided into chunks of length $k$, 
$\tu_i=u_{ik+1}^{ik+k}\in\calU^k$, 
$i=0,1,2,\ldots$, which are fed into a stochastic finite--state encoder, defined by the following equations:

\begin{align}
	\mbox{Pr}\{\tX_i=\tx|\tu_i=\tu,~s_i^{\mbox{\tiny e}}=s\}&=P(\tx|\tu,s),~~~~i=0,1,2,\ldots\\
	s_{i+1}^{\mbox{\tiny e}}&=h(\tu_i,s_i^{\mbox{\tiny e}}),~~~~i=0,1,2,\ldots,
\end{align}
where the variables of these equations are defined as follows: $\tX_i$ is a random vector taking values in $\calX^m$, $\calX$ being the 
$\beta$-ary input alphabet of the channel and $m$ being a positive integer,
$\tx\in\calX^m$ is a realization of $\tX_i$, $s_{i}^{\mbox{\tiny e}}$ is the state of the encoder at time $i$, which takes on
values in a finite set of states, $\calS^{\mbox{\tiny e}}$, of size 
$q_{\mbox{\tiny e}}$. The variable $\tu$ is an arbitrary member of $\calU^k$. The function $h:\calU^k\times\calS^{\mbox{\tiny e}}\to\calS^{\mbox{\tiny e}}$
is called the {\em next-state function} of the encoder.\footnote{More generally, we could have defined both $s_{i+1}^{\mbox{\tiny e}}$
and $\tx_i$ to be random functions of the $(\tu_i,s_i^{\mbox{\tiny e}})$ by a conditional joint distribution,
$\mbox{Pr}\{\tX_i=\tx,s_{i+1}^{\mbox{\tiny e}}=s|\tu_i=\tu,~s_i^{\mbox{\tiny e}}=s'\}$. However, it makes sense to let the encoder
state sequence evolve deterministically in response to the input $\bu$ since the state designates 
the memory of the encoder to past inputs.}
Finally, $P(\tx|\tu,s)$, $\tu\in\calU^k$, $s\in\calS^{\mbox{\tiny e}}$, $\tx\in\calX^m$, is a conditional probability distribution
function, i.e., $\{P(\tx|\tu,s)\}$ are all non-negative and
$\sum_{\tx}P(\tx|\tu,s)=1$ for all $(\tu,s)\in\calU^k\times\calS^{\mbox{\tiny e}}$. 
Without loss of generality, we assume that the initial state of the encoder, $s_0^{\mbox{\tiny e}}$, 
is some fixed member of $\calS^{\mbox{\tiny e}}$.
The ratio

\begin{equation}
	\lambda\dfn\frac{m}{k}
\end{equation}
is referred to as the {\em bandwidth expansion factor}. It should be pointed out that the parameters $k$ and $m$ are fixed integers,
which are not necessarily large (e.g., $k=2$ and $m=3$ are valid values of $k$ and $m$).
The concatenation of output vectors from the encoder, $\tx_0,\tx_1,\ldots$, is viewed as a sequence chunks of channel input symbols,
$x_1,x_2,\ldots$, with $\tx_i=x_{im+1}^{im+m}$, similarly as in the above defined partition of the source sequence.

The sequence of encoder outputs, $x_1,x_2,\ldots$, is fed into a discrete memoryless channel (DMC),
henceforth referred to as the {\em main channel},
whose corresponding outputs, $y_1,y_2,\ldots$, are generated according to:

\begin{equation}
	\mbox{Pr}\{Y^N=y^N|X^N=x^N\}=Q_{\mbox{\tiny M}}(y^N|x^N)=\prod_{i=1}^NQ_{\mbox{\tiny M}}(y_i|x_i),
\end{equation}
for every positive integer $N$ and every $x^N\in\calX^N$ and $y^N\in\calY^N$. 
The channel output symbols, $\{y_i\}$, take values in a finite alphabet, $\calY$, of size $\gamma$.

The sequence of channel outputs, $y_1,y_2,\ldots$, is divided into chunks of length $m$, $\ty_i=y_{im+1}^{im+m}$,
$i=0,1,2,\ldots$, which are fed into a deterministic finite--state decoder, defined according to the following recursive
equations:

\begin{eqnarray}
	\tv_i&=&f(\ty_i,s_i^{\mbox{\tiny d}})\\
	s_{i+1}^{\mbox{\tiny d}}&=&g(\ty_i,s_i^{\mbox{\tiny d}}),
\end{eqnarray}
where the variables in the equations are defined as follows: $\{s_i^{\mbox{\tiny d}}\}$ is the sequence of states of the decoder,
which takes values in a finite set, $\calS^{\mbox{\tiny d}}$ of size $q_{\mbox{\tiny d}}$. The variable $\tv_i\in\calU^k$ is the $i$-th
chunk of $k$ source reconstruction symbols, i.e., $\tv_i=v_{ik+1}^{ik+k}$, $i=0,1,\ldots$. 
The function $f:\calY^m\times\calS^{\mbox{\tiny d}}\to\calU^k$ is called the {\em output function} of the decoder and the function
$g:\calY^m\times\calS^{\mbox{\tiny d}}\to\calS^{\mbox{\tiny d}}$ is the next--state function of the decoder.
The concatenation of the decoder output vectors,
$\tv_0,\tv_1,\ldots$, forms the entire stream of reconstruction symbols, $v_1,v_2,\ldots$.

The output of the main channel, $y_1,y_2,\ldots$, is fed into another DMC, henceforth referred to as the {\em wiretap channel}, 
which generates in response, a corresponding
sequence, $z_1,z_2,\ldots$, according to

\begin{equation}
	\mbox{Pr}\{Z^N=z^N|Y^N=y^N\}=Q_{\mbox{\tiny W}}(z^N|y^N)=\prod_{i=1}^NQ_{\mbox{\tiny W}}(z_i|y_i),
\end{equation}
where $\{Z_i\}$ and $\{z_i\}$ take values in a finite alphabet $\calZ$. We denote the cascade of channels $Q_M$ and $Q_W$ by $Q_{MW}$,
that is

\begin{equation}
	Q_{\mbox{\tiny MW}}(z|x)=\sum_{y\in\calY}Q_{\mbox{\tiny M}}(y|x)Q_{\mbox{\tiny W}}(z|y).
\end{equation}

We seek a communication system $(P,h,f,g)$ which satisfies two requirements:
\begin{enumerate}
\item For a given $\epsilon_{\mbox{\tiny r}} > 0$, the system satisfies the following reliability requirement:
The bit error probability is guaranteed to be less than $\epsilon_{\mbox{\tiny r}}$, i.e.,

\begin{equation}
	\label{reliability}
	P_{\mbox{\tiny b}}\dfn\frac{1}{k}\sum_{i=1}^k\mbox{Pr}\{V_i\ne u_i\}\le\epsilon_{\mbox{\tiny r}}
\end{equation}
for every $(u_1,\ldots,u_k)$ and every combination of initial states of the encoder and the decoder, where $\mbox{Pr}\{\cdot\}$
is defined w.r.t.\ the randomness of the encoder and the main channel.

\item For a given $\epsilon_{\mbox{\tiny s}} > 0$, the system satisfies the following security 
requirement: For every sufficiently large positive integer $n$,

\begin{equation}
\label{security}
\max_{\mu} I_\mu(U^n;Z^N)\le n\epsilon_{\mbox{\tiny s}},
\end{equation}
where $N=n\lambda$ and $I_\mu(U^n;Z^N)$ is the mutual information between $U^n$ and $Z^N$, induced by
an input distribution $\mu=\{\mu(u^n),~u^n\in\calU^n\}$ and the system, $\{P(z^N|u^n),~u^n\in\calU^n,
z^N\in\calZ^N\}$.
\end{enumerate}

As for the reliability requirement, note that the larger is $k$, the requirement becomes less stringent.
Concerning the security requirement, ideally, we would like to have perfect secrecy, which means that 
$P(z^N|u^n)$ would be independent of $u^n$ (see also \cite{me13}),
but it is more realistic to allow a small deviation from this idealization.
This security metric is actually the maximum mutual information metric, or equivalently (see \cite{BTV12}) the semantic security, as mentioned in the Introduction.

\section{Results}
\label{results}

We begin with definitions of two more quantities. The first is the {\em secrecy capacity} \cite{Wyner75}, \cite{BB11}, which is
the supremum of all coding rates for which there exist block codes that maintain both an arbitrarily small error probability at the legitimate decoder
and an equivocation arbitrarily close to the unconditional entropy of the source. The secrecy capacity is given by

\begin{equation}
C_{\mbox{\tiny s}}=\max_{P_X} I(X;Y|Z)=\max_{P_X}[I(X;Y)-I(X;Z)],
\end{equation}
with $P_{XYZ}(x,y,z)=P_X(x)\times Q_{\mbox{\tiny M}}(y|x)Q_{\mbox{\tiny W}}(z|y)$ for all $(x,y,z)\in\calX\times\calY\times\calZ$.

The second quantity we need to define is the LZ complexity \cite{ZL78}. 
Consider the process of {\em incremental parsing} the source vector, $u^n$, that is, sequentially parsing
this sequence into distinct phrases, such that each new parsed phrase is the shortest string that has not been obtained before as a phrase, with a possible
exception of the last phrase, which might be incomplete. Let $c(u^n)$ denote the number of resulting phrases. For example, if $n=10$ and
$u^{10}=(0000110110)$ then incremental parsing (from left to right) yields $(0,00,01,1,011,0)$ and so, $c(u^{10})=6$. We define the {\em LZ complexity}
of the individual sequence, $u^n$, as 

\begin{equation}
	\rho_{\mbox{\tiny LZ}}(u^n)\dfn\frac{c(u^n)\log c(u^n)}{n}.
\end{equation}
As was shown by Ziv and Lempel in their seminal paper \cite{ZL78}, for large $n$, the LZ complexity, 
$\rho_{\mbox{\tiny LZ}}(u^n)$, is essentially the best compression ratio that can be achieved by any information lossless, finite--state encoder (up to some
negligibly small terms, for large $n$), and it can be viewed as the individual--sequence analogue of the entropy rate.

Before moving on to present our first main result, a simple comment is in order.
Even in the traditional probabilistic setting, given a source with entropy $H$ and a channel with capacity $C$, reliable communication cannot be
accomplished unless $H\le\lambda C$, where $\lambda$ is the bandwidth expansion factor. Since both $H$ and $C$ are given and only $\lambda$ is
a under the control of the system designer, it is natural to state this condition as a lower bound to bandwidth expansion factor, i.e.,
$\lambda \ge H/C$.
By the same token, in the presence of a secrecy constraint, $\lambda$ must not fall below $H/C_{\mbox{\tiny s}}$.
Our converse theorems for individual sequences will be presented in the same spirit, where the entropy $H$ at the numerator 
will be replaced by an expression whose main term is 
the Lempel-Ziv compressibility.

We assume, without essential loss of generality, that $k$ divides $n$ (otherwise, omit the last $(n \mod k)$ symbols of $u^n$ and replace $n$ by
$k\cdot\lfloor n/k\rfloor$ without affecting the asymptotic behavior as $n\to\infty$).
Our first main result is the following:
\begin{theorem}
	\label{thm1}
Consider the problem setting defined in Section \ref{ndps}. If there exists a stochastic encoder with $q_{\mbox{\tiny e}}$ states and
a decoder with $q_{\mbox{\tiny d}}$ states that together satisfy the reliability constraint (\ref{reliability}) and the
security constraint (\ref{security}), then the bandwidth expansion factor $\lambda$ must be lower bounded as follows.

\begin{equation}
\lambda\ge\frac{\rho_{\mbox{\tiny LZ}}(u^n)-\Delta(\epsilon_{\mbox{\tiny r}})-
\epsilon_{\mbox{\tiny s}}-\zeta_n(q_{\mbox{\tiny d}},k)}{C_{\mbox{\tiny s}}},
\end{equation}
where 

\begin{equation}
\Delta(\epsilon_{\mbox{\tiny r}})\dfn h_2(\epsilon_{\mbox{\tiny r}})+\epsilon_{\mbox{\tiny r}}\cdot\log(\alpha-1),
\end{equation}
with $h_2(\epsilon_{\mbox{\tiny r}})=-\epsilon_{\mbox{\tiny r}}\log \epsilon_{\mbox{\tiny r}}-
(1-\epsilon_{\mbox{\tiny r}})\log (1-\epsilon_{\mbox{\tiny r}})$ being the binary entropy function, 
and

\begin{equation}
	\zeta_n(q_{\mbox{\tiny d}},k)=\min_{\{\ell~\mbox{divides}~n/k\}}\bigg[\frac{\log q_{\mbox{\tiny d}}+1}{k\ell}+\frac{2k\ell(\log\alpha+1)^2}
	{(1-\epsilon_n)\log n}+\frac{2k\ell\alpha^{2k\ell}\log\alpha}{n}\bigg],
\end{equation}
with $\epsilon_n\to 0$ as $n\to\infty$.
\end{theorem}
The proof of Theorem \ref{thm1}, like all other proofs in this article, is deferred to Section \ref{proofs}.

\vspace{0.15cm}

\noindent
{\bf Discussion.}
A few comments are in order with regard to Theorem \ref{thm1}.\\

\noindent
1. {\em Irrelevance of $q_{\mbox{\tiny e}}$.} 
It is interesting to note that as far as the encoding and decoding resources are concerned, the lower bound depends on $k$ and $q_{\mbox{\tiny d}}$, but not on
the number of states of the encoder, $q_{\mbox{\tiny e}}$. This means that the same lower bound continues to hold even if the encoder has an unlimited
number of states. Pushing this to the extreme, even if the encoder has room to store the entire past, the lower bound 
of Theorem \ref{thm1} would remain unaltered. The crucial
bottleneck is therefore in the finite memory resources associated with the decoder, where the memory may help to reconstruct the source by exploiting empirical
dependencies with the past. The dependence on $q_{\mbox{\tiny e}}$, however, will appear later, when we discuss local randomness resources as well as in the
extension to the case of decoder side information.\\

\noindent
2. {\em The redundancy term $\zeta_n(q_{\mbox{\tiny d}},k)$.} 
A technical comment is in order concerning the term $\zeta_n(q_{\mbox{\tiny d}},k)$, which involves minimization over all divisors of $n/k$, where we have
already assumed that $n/k$ is integer. Strictly speaking, if $n/k$ happens to be a prime, this minimization is not very meaningful as
$\zeta_n(q_{\mbox{\tiny d}},k)$ would be relatively large. If this the case, a better bound will be obtained if one omits some of the last symbols of $u^n$
and thereby reduce $n$, to say, $n'$, so that $n'/k$ has a richer set of factors. 
Consider, for example, the choice $\ell=\ell_n=\lfloor \sqrt{\log n}\rfloor$ (instead of minimizing over $\ell$) 
and replace $n/k$ by the $n/k-(n/k~\mod~\ell_n)$, without essential loss of tightness.
This way, $\zeta_n(q_{\mbox{\tiny d}},k)$ would tend to zero as $n\to\infty$, for fixed $k$ and $q_{\mbox{\tiny d}}$.\\

\noindent
3. {\em Achievability.} Having established that $\zeta_n(q_{\mbox{\tiny d}},k)\to 0$, 
and given that $\epsilon_{\mbox{\tiny r}}$ and $\epsilon_{\mbox{\tiny s}}$ are small,
it is clear that the main term at the numerator of the lower bound of 
Theorem \ref{thm1} is the term $\rho_{\mbox{\tiny LZ}}(u^n)$, which is, as mentioned
earlier, the individual--sequence analogue of the entropy of the source \cite{ZL78}. In other words, $\lambda$ cannot be much smaller than
$\lambda_{\mbox{\tiny L}}(u^n)=\rho_{\mbox{\tiny LZ}}(u^n)/C_{\mbox{\tiny s}}$. A matching achievability scheme would most naturally be based on separation:
first apply variable--rate
compression\footnote{Note that in this individual--sequence setting, the distinction between fixed--rate codes and variable--rate codes is not
quite meaningful to begin with, because there is only one source sequence, $u^n$, to handle. The best reference encoder--decoder pair
depend on $u^n$, and hence so it is rate.}
of $u^n$ to about $n\rho_{\mbox{\tiny LZ}}(u^n)$ bits using the LZ algorithm \cite{ZL78}, and then feed the resulting compressed
bit-stream into a good code for the wiretap channel \cite{Wyner75} with codewords of length about 

\begin{equation}
N=n\lambda_{\mbox{\tiny L}}(u^n)\sim
	\frac{n\rho_{\mbox{\tiny LZ}}(u^n)}{C_{\mbox{\tiny s}}(1-\delta)}, 
\end{equation}
where $\delta$ is an arbitrarily small (but positive) margin to keep the coding rate strictly smaller than 
$C_{\mbox{\tiny s}}$. But to this end, the decoder must know $N$. One possible solution is
that before the actual encoding of each $u^n$, one
would use a separate, auxiliary fixed code that encodes the value of the number of compressed bits, $n\rho_{\mbox{\tiny LZ}}(u^n)$, 
using $\log(n\log\alpha)$ bits
(as $n\log\alpha$ is about the number of possible values that $n\rho_{\mbox{\tiny LZ}}(u^n)$ can take)
and protect it using a channel code of rate less than $C_{\mbox{\tiny s}}(1-\delta)$.
Since the length of this auxiliary code grows only
logarithmically with $n$ (as opposed to the `linear' growth of $n\rho_{\mbox{\tiny LZ}}(u^n)$),
the overhead in using the auxiliary code is asymptotically negligible. The auxiliary code and the main code will be used alternately,
first the auxiliary code, and then the main code for each $n$-tuple of the source. The main channel code is actually an array of
codes, one for each possible value of $n\rho_{\mbox{\tiny LZ}}(u^n)$. Once the auxiliary decoder has decoded this number, the
corresponding main decoder is used. Overall, the resulting bandwidth expansion factor is about

\begin{equation}
        \lambda\approx \frac{n\rho_{\mbox{\tiny LZ}}(u^n)+\log(n\log\alpha)}{nC_{\mbox{\tiny s}}(1-\delta)}
        =\frac{\rho_{\mbox{\tiny LZ}}(u^n)}
        {C_{\mbox{\tiny s}}(1-\delta)}+O\left(\frac{\log n}{n}\right).
\end{equation}
Another, perhaps simpler and better, approach is to use the LZ
algorithm in the mode of a variable-to--fixed length code: Let the the length of the channel codeword,
$N$, be fixed, and start to compress $\bu=(u_1,u_2,\ldots)$
until obtaining $n\rho_{\mbox{\tiny LZ}}(u^n)=N\cdot C_{\mbox{\tiny s}}(1-\delta)$ compressed
bits. Then,

\begin{equation}
        \lambda=\frac{N}{n}=\frac{\rho_{\mbox{\tiny LZ}}(u^n)}{C_{\mbox{\tiny s}}(1-\delta)}.
\end{equation}
Of course, these coding schemes require decoder memory that grows exponentially in $n$, and not just a fixed number, $q_{\mbox{\tiny d}}$, and therefore
strictly speaking, there is a gap between the achievability and the converse result of Theorem 2. 
However, this gap is closed asymptotically, once we take the limit of $q_{\mbox{\tiny d}}\to 0$ after the limit $n\to\infty$, and we consider successive
application of these codes over many blocks. The same approach appears also in \cite{Ziv78}, \cite{Ziv80}, \cite{Ziv84}, \cite{ZL78}, as well as in later 
related work.\\

\noindent
This concludes the discussion on Theorem \ref{thm1}.$\Box$\\

We next focus on local randomness resources that are necessary when the full secrecy capacity is exploited.
Specifically, suppose that the stochastic encoder $\{P(\tx|\tu,s),~\tx\in\calX^n,~\tu\in\calU^k,~s\in\calS^{\mbox{\tiny e}}\}$ is implemented
as a deterministic encoder with an additional input of purely random bits, i.e.,

\begin{equation}
	\label{explicitrandomization}
        \tx_i=a(\tu_i,s_i^{\mbox{\tiny e}},\tb_i),
\end{equation}
where $\tb_i=b_{ij+1}^{ij+j}$
is a string of $j$ purely random bits. The question is the following: How large must $j$ be in order to achieve full secrecy?
Equivalently, what is the minimum necessary rate of random bits for local randomness at the encoder for secure coding at the maximum reliable rate?
In fact, this question may be interesting on its own right, regardless of the individual--sequence setting and finite--state
encoders and decoders, but even for ordinary block coding (which is the special case of 
$q_{\mbox{\tiny e}}=q_{\mbox{\tiny d}}=1$) and in the traditional probabilistic setting.
The following theorem answers this question.

\begin{theorem}
\label{thm2}
Consider the problem setting defined in Section \ref{ndps} and let $\lambda$ meet the lower bound of Theorem 1.
If there exists an encoder (\ref{explicitrandomization}) with $q_{\mbox{\tiny e}}$ states and a decoder
with $q_{\mbox{\tiny d}}$ states that jointly satisfy the reliability constraint (\ref{reliability}) and
the security constraint (\ref{security}), then

\begin{equation}
j\ge mI(X^*;Z^*)- k\epsilon_{\mbox{\tiny s}}-\frac{\log q_{\mbox{\tiny e}}}{\ell}
\end{equation}
where $X^*$ is the random variable that achieves $C_{\mbox{\tiny s}}$ and
$\ell$ is the achiever of $\zeta_n(q_{\mbox{\tiny d}},k)$.
\end{theorem}

Note that the lower bound of Theorem \ref{thm2} 
depends on $q_{\mbox{\tiny e}}$, as opposed to Theorem \ref{thm1}, where it depended only on $q_{\mbox{\tiny d}}$.
Since $\epsilon_{\mbox{\tiny s}}$ is assumed small and $\ell\to\infty$, it is clear that main term is $mI(X^*;Z^*)$, i.e.,
the bit rate must be essentially at least as large as $I(X^*;Z^*)$ random bits per channel use, or equivalently,
$\lambda I(X^*;Z^*)$ bits per source symbol. It is interesting to note that Wyner's code \cite{Wyner75} asymptotically achieves this bound
when the coding rate saturates the secrecy capacity, because the subcode that can be decoded by the wiretapper (within each given bin) is of rate
about $I(X^*;Z^*)$, and it encodes just the bits of the local randomness. So when working at the full secrecy capacity, Wyner's code is optimal, not
only in terms of the optimal trade-off between reliability and security, but also in terms of minimum consumption of local, purely random bits.

\section{Side Information at the Decoder with Partial Leakage to the Wiretapper}
\label{si}

Consider next an extension of our model to the case where there are
side information sequences, $w^n=(w_1,\ldots,w_n)$ and $\dw^n=(\dw_1,\ldots,\dw_n)$,
available to the decoder and the wiretapper, respectively. For the purpose of a converse theorem, we assume that $w^n$ is available to the encoder too, whereas in
the achievability part, we will comment also on the case where it is not.
We will assume that $w^n$ is a deterministic sequence, but $\dw^n$ is a realization of a random vector
$\dW^n=(\dW_1,\ldots,\dW_n)$, which is a noisy version of $w^n$. In other words, it is generated from
$w^n$ by another memoryless channel, $Q_{\dW^n|W^n}(\dw^n|w^n)=\prod_{i=1}^nQ_{\dW|W}(\dw_i|w_i)$.
The symbols of $\{w_i\}$ and $\{\dw_i\}$ take values in finite alphabets, $\calW$ and $\dot{\calW}$, respectively.
There are two extreme
important special cases: (i) $\dW^n=w^n$ almost surely, which is the case of totally insecure side information that fully leaks to the wiretapper, and (ii)
$\dW^n$ is degenerated (or independent of $w^n$), which is the case of secure side information with no leakage to the wiretapper. Every intermediate situation between
these two extremes is a situation of partial leakage.
The finite--state encoder model is now re-defined according to:

\begin{eqnarray}
        \mbox{Pr}\{\tX_i=\tx|\tu_i=\tu,~\tw_i=\tw,s_i^{\mbox{\tiny e}}=s\}&=&P(\tx|\tu,\tw,s),~~~~i=0,1,2,\ldots\\
        s_{i+1}^{\mbox{\tiny e}}&=&h(\tu_i,\tw_i,s_i^{\mbox{\tiny e}}),~~~~i=0,1,2,\ldots,
\end{eqnarray}
where $\tw_i=w_{ik+1}^{ik+k}$, $i=0,1,\ldots,n/k-1$. Likewise, the decoder is given by
\begin{eqnarray}
        \tv_i&=&f(\ty_i,\tw_i,s_i^{\mbox{\tiny d}})\\
        s_{i+1}^{\mbox{\tiny d}}&=&g(\ty_i,\tw_i,s_i^{\mbox{\tiny d}}),
\end{eqnarray}
and the wiretapper has access to $Z^N$ and $\dW^n$. 
Accordingly, the security constraint is modified as follows: For a given $\epsilon_{\mbox{\tiny s}} > 0$
and for every sufficiently large $n$,

\begin{equation}
\label{security2}
\max_{\mu} I_\mu(U^n;Z^N|\dW^n)\le n\epsilon_{\mbox{\tiny s}},
\end{equation}
where $I_\mu(U^n;Z^N|\dW^n)$ is the conditional mutual information between $U^n$ and $Z^N$ given $\dW^n$, induced by
$\mu=\{\mu(u^n,\dw^n),~u^n\in\calU^n,~\dw^n\in\dot{\calW}^n\}$ and the system, $\{P(z^N|u^n),~u^n\in\calU^n,
z^N\in\calZ^N\}$, where $\mu(u^n,\dw^n)=\sum_{w^n}\mu(u^n,w^n)Q_{\dW^n|W^n}(\dw^n|w^n)$.

In order to present the extension of Theorem \ref{thm1} to incorporate side information, we first need to define the
extension of the LZ complexity to include side information, namely, to define the conditional LZ complexity (see also \cite{Ziv85}).
Given $u^n$ and $w^n$,
let us apply the incremental
parsing procedure of the LZ algorithm
to the sequence of pairs $((u_1,w_1),(u_2,w_2),\ldots,(u_n,w_n))$.
According to this procedure, all phrases are distinct
with a possible exception of the last phrase, which might be incomplete.
Let $c(u^n,w^n)$ denote the number of distinct phrases.
For example,\footnote{The same example appears in \cite{Ziv85}.} if

\begin{eqnarray}
u^6&=&0~|~1~|~0~0~|~0~1|\nonumber\\
w^6&=&0~|~1~|~0~1~|~0~1|\nonumber
\end{eqnarray}
then $c(u^6,w^6)=4$.
Let $c(w^n)$ denote the resulting number of distinct phrases
of $w^n$, and let $w(l)$ denote the $l$-th distinct $w$--phrase,
$l=1,2,...,c(w^n)$. In the above example, $c(w^6)=3$. Denote by
$c_l(u^n|w^n)$ the number of occurrences of $w(l)$ in the
parsing of $w^n$, or equivalently, the number of distinct $u$-phrases
that jointly appear with $w(l)$. Clearly, $\sum_{l=1}^{c(w^n)} c_l(u^n|w^n)=
c(u^n,w^n)$. In the above example, $w(1)=0$, $w(2)=1$, $w(3)=01$,
$c_1(u^6|w^6)=c_2(u^6|w^6)=1$, and $c_3(u^6|w^6)=2$. Now, the conditional LZ
complexity of $u^n$ given $w^n$ is defined as

\begin{equation}
\rho_{LZ}(u^n|w^n)\dfn\frac{1}{n}\sum_{l=1}^{c(w^n)}c_l(u^n|w^n)\log c_l(u^n|w^n).
\end{equation}

We are now ready to present the main result of this section.
\begin{theorem}
\label{thm3}
Consider the problem setting defined in Section \ref{ndps} along with the above--mentioned modifications to incorporate side information.
If there exists a stochastic encoder with $q_{\mbox{\tiny e}}$ states and
a decoder with $q_{\mbox{\tiny d}}$ states that together satisfy the reliability constraint (\ref{reliability}) and the
security constraint (\ref{security2}), then its bandwidth expansion factor $\lambda$ must be lower bounded as follows.

\begin{equation}
\lambda\ge\frac{\rho_{\mbox{\tiny LZ}}(u^n|w^n)-\Delta(\epsilon_{\mbox{\tiny r}})-
	\epsilon_{\mbox{\tiny s}}-\eta_n(q_{\mbox{\tiny e}}\cdot q_{\mbox{\tiny d}},k)}{C_{\mbox{\tiny s}}},
\end{equation}
where

\begin{equation}
	\eta_n(q_{\mbox{\tiny e}}\cdot q_{\mbox{\tiny d}},k)=
	\min_{\{\ell~\mbox{divides}~n/k\}}\bigg[\frac{\log(q_{\mbox{\tiny d}}q_{\mbox{\tiny e}})+1}{k\ell}+\frac{\log(4A^2)}
	{(1-\epsilon_n)\log n}+\frac{A^2\log(4A^2)}{n}\bigg],
\end{equation}
with $\epsilon_n\to 0$ as $n\to\infty$ and $A=[(\alpha\omega)^{k\ell+1}-1]/[\alpha\omega-1]$, $\omega$ being the size of $\calW$.
\end{theorem}

Note that the lower bound of Theorem \ref{thm3} does not depend on the noisy side information at the wiretapper or on the channel $Q_{\dW|W}$ that generates it
from $w^n$. It depends only on $u^n$ and $w^n$ in terms of the data available in the system. 
Clearly, as it is a converse theorem, if it allows the side information to be available also at the encoder, then it definitely applies also to the case
where the encoder does not have access to $w^n$.
Interestingly, the encoder and the legitimate decoder
act as if the wiretapper had the {\em clean} side information, $w^n$. While it is quite obvious that protection 
against availability of $w^n$ at the wiretapper is sufficient 
for protection against availability of $\dW^n$ (as $\dW^n$ is a degraded version of $w^n$), 
it is not quite trivial that this should be also {\em necessary}, as the above converse theorem asserts.
It is also interesting to note that here, the bound depends also on $q_{\mbox{\tiny e}}$, and not only $q_{\mbox{\tiny d}}$, as in Theorem \ref{thm1}.
However, this dependence on $q_{\mbox{\tiny e}}$ disappears in the special case where $\dW^n=w^n$ with probability one.

We next discuss the achievability of the lower bound of Theorem \ref{thm3}. If encoder has access to $w^n$, then the first step would be to
apply the conditional LZ algorithm (see \cite[proof of Lemma 2]{Ziv85}, \cite{UK03}), thus compressing $u^n$ 
to about $n\rho_{\mbox{\tiny LZ}}(u^n|w^n)$ bits, and the second step would be good channel coding for the wiretap channel,
using the same methods as described in the previous section. If, however, the encoder does not have access 
to $w^n$, the channel coding part is still as before, but 
the situation with the source coding part 
is somewhat more involved, since neither the encoder nor the decoder can calculate the target bit rate, $\rho_{\mbox{\tiny LZ}}(u^n|w^n)$, as neither
party has access to both $u^n$ and $w^n$. However, this source coding rate can essentially be achieved,
provided that there is a low--rate noiseless feedback channel
from the legitimate decoder to the encoder. The following scheme is in the spirit of the one proposed by Draper \cite{Draper04}, but with a few
modifications.

The encoder implements random binning for all source sequences in $\calU^n$, that is, for each member of $\calU^n$ an index 
is drawn independently, under the uniform distribution over $\{0,1,2,\ldots,\alpha^n-1\}$, which is represented by 
its binary expansion, $b(u^n)$, of length $n\log\alpha$ bits. We select a large positive integer $r$, but keep $r\ll n$ (say, $r=\sqrt{n}$ or $r=\log^2n$).
The encoder transmits the bits of $b(u^n)$ incrementally, $r$ bits at a time, until it receives from the decoder {\tt ACK}. 
Each chunk of $r$ bits is fed into a good channel code for the wiretap channel, 
at a rate slightly less than $C_{\mbox{\tiny s}}$. 
At the decoder side, this channel code is decoded (correctly, with high probability, for large $r$).
Then, for each $i$ ($i=1,2,\ldots$), after having decoded
the $i$-th chunk of $r$ bits of $b(u^n)$, the decoder
creates the list $\calA_i(u^n)=\{\dot{u}^n:~
[b(\dot{u}^n)]^{ir}=[b(u^n)]^{ir}\}$, where $[b(\dot{u}^n)]^l$ denotes the string formed by the first $l$ bits of $b(\dot{u}^n)$.
For each $\dot{u}^n\in\calA_i(u^n)$, the decoder calculates $\rho_{\mbox{\tiny LZ}}(\dot{u}^n|w^n)$. 
Fix an arbitrarily small $\delta > 0$, which controls the trade-off between error probability and compression rate. If 
$n\rho_{\mbox{\tiny LZ}}(\dot{u}^n|w^n)\le i\cdot r-n\delta$ for some $\dot{u}^n\in\calA_i(u^n)$, the decoder sends {\tt ACK} on the feedback channel
and outputs the reconstruction, $\dot{u}^n$, with the smallest $\rho_{\mbox{\tiny LZ}}(\dot{u}^n|w^n)$ among all members of $\calA_i(u^n)$.
If no member of $\calA_i(u^n)$ satisfies $n\rho_{\mbox{\tiny LZ}}(\dot{u}^n|w^n)\le i\cdot r-n\delta$, 
the receiver waits for the next chunk of $r$ compressed bits,
and it does not send {\tt ACK}. The probability of source-coding error after the $i$-th chunk is upper bounded by

\begin{eqnarray}
	P_{\mbox{\tiny e}}(i)&\lea&|\{\dot{u}^n\ne u^n:~n\rho_{\mbox{\tiny LZ}}(\dot{u}^n|w^n)\le i\cdot r-n\delta\}|\cdot 2^{-i\cdot r}\nonumber\\
	&\leb&\exp_2\left\{i\cdot r-n\delta+O\left(\frac{\log(\log n)}{\log n}\right)\right\}\cdot 2^{-i\cdot r}\nonumber\\
	&=&\exp_2\left\{-n\delta+O\left(\frac{\log(\log n)}{\log n}\right)\right\}\nonumber\\
	&\to& 0~~~~\mbox{as}~n\to\infty,
\end{eqnarray}
where in (a), the factor $2^{-i\cdot r}$ is the probability that $[b(\dot{u}^n)]^{ir}=[b(u^n)]^{ir}$ for each member of the set
$\{\dot{u}^n\ne u^n:~n\rho_{\mbox{\tiny LZ}}(\dot{u}^n|w^n)\le i\cdot r-n\delta\}$ and (b) is based on \cite[eq.\ (A.13)]{Ziv85}.
Clearly, it is guaranteed that an {\tt ACK} will be received at the encoder (and hence the transmission will stop),
no later than after the transmission of chunk no.\ 
$i^*$, where $i^*$ is the smallest integer $i$ such that $i\cdot r\ge n\rho_{\mbox{\tiny LZ}}(u^n|w^n)+n\delta$, namely,
$i^*=\lceil[n\rho_{\mbox{\tiny LZ}}(u^n|w^n)+n\delta]/r\rceil$,
which is the stage at which at least the correct source sequence
begins to satisfy the condition $n\rho_{\mbox{\tiny LZ}}(u^n|w^n)\le i\cdot r-n\delta$. 
Therefore, the compression ratio is no worse than $i^*\cdot r/n=\lceil n[\rho_{\mbox{\tiny LZ}}(u^n|w^n)+\delta]/r\rceil\cdot r/n\le 
\rho_{\mbox{\tiny LZ}}(u^n|w^n)+\delta+r/n$.
The overall probability of source--coding error is then
upper bounded by

\begin{equation}
	P_{\mbox{\tiny e}}=\mbox{Pr}\bigcup_{i=1}^{i^*}\{\mbox{error at state}~i\}\le\sum_{i=1}^{i^*} P_{\mbox{\tiny e}}(i)\le
	\left(\frac{n\log\alpha}{r}+1\right)\cdot\exp_2\left\{-n\delta+O\left(\frac{\log(\log n)}{\log n}\right)\right\},
\end{equation}
which still tends to zero as $n\to\infty$. As for channel--coding errors, the probability that at least one chunk will
be decoded incorrectly is upper bounded by $(\frac{n\log\alpha}{r}+1)\cdot e^{-rE}$, where $E$ is an achievable error exponent of channel coding at
the given rate. Thus, if $r$ grows at any rate faster than logarithmic, but sub-linear in $n$, then the overall channel--coding 
error probability tends to zero and, at the same time, the compression redundancy, $r/n$, tends to zero too.

To show that the security constraint (\ref{security2}) is satisfied too, consider an arbitrary assignment $\mu$ of random vectors $(U^n,W^n)$ and
let us denote by $B$ the string of $I(X^N;Z^N)-N\epsilon$ bits of local randomness in Wyner's code \cite{Wyner75}.
Then,

\begin{eqnarray}
        I(X^N;Z^N)&=&H(Z^N)-H(Z^N|X^N)\nonumber\\
        &\gea&H(Z^N)-H(Z^N|U^n,B)\nonumber\\
        &\geb&H(Z^N|\dW^n)-H(Z^N|U^n,B)\nonumber\\
        &\eqc&H(Z^N|\dW^n)-H(Z^N|U^n,B,\dW^n)\nonumber\\
        &=&I(U^n,B;Z^N|\dW^n)\nonumber\\
        &=&H(U^n,B|\dW^n)-H(U^n,B|Z^n,\dW^n)\nonumber\\
        &=&H(U^n|\dW^n)+H(B|U^n,\dW^n)-H(U^n|Z^N,\dW^n)-H(B|Z^N,\dW^n,U^n)\nonumber\\
        &\eqd&H(U^n|\dW^n)+H(B)-H(U^n|Z^N,\dW^n)-H(B|Z^N,\dW^n,U^n)\nonumber\\
        &\gee&H(U^n|\dW^n)+H(B)-H(U^n|Z^N,\dW^n)-H(B|Z^N,U^n)\nonumber\\
        &\gef&H(U^n|\dW^n)+[I(X^N;Z^N)-N\epsilon]-H(U^n|Z^N,\dW^n)-n\delta_n\nonumber\\
        &=&I(X^N;Z^N)+I_\mu(U^n;Z^N|\dW^n)-n(\lambda\epsilon+\delta_m),
\end{eqnarray}
where (a) is since $(U^n,B)\to X^N\to Z^N$ is a Markov chain,
(b) is since conditioning reduces entropy,
(c) is since $\dW^n\to (U^n,B)\to Z^N$ is a Markov chain,
(d) is since $B$ is independent of $(U^n,\dW^n)$,
(e) is since conditioning reduces entropy, and
(f) is since in Wyner coding, $B$ can be reliably decoded given $(Z^N,U^n)$ ($\delta_n$ is understood to be small, and 
recall that $W^n$ is not needed in the channel decoding phase, but
only in the Slepian--Wold decoding phase),
and that the length of $B$ is chosen to be $I(X^N;Z^N)-N\epsilon$.
Comparing the right-most side to the left-most side, we readily obtain:

\begin{equation}
        I_\mu(U^n;Z^N|\dW^n)\le n(\lambda\epsilon+\delta_n),
\end{equation}
which can be made arbitrarily small.

\section{Proofs}
\label{proofs}

We begin this section by establishing more notation conventions to be used throughout all proofs.

Let $n \gg k$ be a positive integer and let $\ell$ be such that $K\dfn\ell\cdot k$ divides $n$. Consider the partition of $u^n$
into $n/K$ non--overlapping blocks of length $K$,

\begin{eqnarray}
	& &(\tu_0,\tu_1,\ldots,\tu_{\ell-1}),
	(\tu_{\ell},\tu_{\ell+1},\ldots,\tu_{2\ell-1}),\ldots,(\tu_{n/k-\ell},\tu_{n/k-\ell+1},\tu_{n/k-1})\nonumber\\
	&=&(u_1^K,u_{K+1}^{2K},\ldots,u_{n-K+1}^n)
\end{eqnarray}
and apply the same partition to $v^n$. The corresponding channel 
input and output sequences are of length $N=n\lambda$.
Let $M=\ell\cdot m=K\lambda$ and consider the parallel partition of the channels input and output sequences according to

\begin{eqnarray}
	 & &(\tx_0,\tx_1,\ldots,\tx_{\ell-1}),
        (\tx_{\ell},\tx_{\ell+1},\ldots,\tx_{2\ell-1}),\ldots,(\tx_{N/m-\ell},\tx_{N/m-\ell+1},\ldots,\tx_{N/m-1})\nonumber\\
	 & &(\ty_0,\ty_1,\ldots,\ty_{\ell-1}),
        (\ty_{\ell},\ty_{\ell+1},\ldots,\ty_{2\ell-1}),\ldots,(\ty_{N/m-\ell},\ty_{N/m-\ell+1},\ldots,\ty_{N/m-1})\nonumber\\
	 & &(\tz_0,\tz_1,\ldots,\tz_{\ell-1}),
        (\tz_{\ell},\tz_{\ell+1},\ldots,\tz_{2\ell-1}),\ldots,(\tz_{N/m-\ell},\tz_{N/m-\ell+1},\ldots,\tz_{N/m-1}).
\end{eqnarray}
For the sake of brevity, we henceforth denote $(\tu_{i\ell},\ldots,\tu_{(i+1)\ell-1})$ by $\tu_{i\ell}^{(i+1)\ell-1}$ and use the
same notation rule for all other sequences. Next, define the joint empirical distribution:

\begin{eqnarray}
        & &P_{\hU^K\hX^M\hY^M\hZ^M\hS^{\mbox{\tiny e}}\hS^{\mbox{\tiny d}}}
        (u^K,x^M,y^M,z^M,s^{\mbox{\tiny e}},s^{\mbox{\tiny d}})=\nonumber\\
        & &\frac{K}{n}\sum_{i=0}^{n/K-1}\delta\{\tu_{i\ell}^{(i+1)\ell-1}=u^K,
        \tx_{i\ell}^{(i+1)\ell-1}=x^M,\ty_{i\ell}^{(i+1)\ell-1}=y^M,\nonumber\\
        & &\tz_{i\ell}^{(i+1)\ell-1}=z^M,s_{i\ell+1}^{\mbox{\tiny e}}=s^{\mbox{\tiny e}},
	s_{i\ell+1}^{\mbox{\tiny d}}=s^{\mbox{\tiny d}}\},
\end{eqnarray}
and

\begin{equation}
	P_{\hU^KX^MY^MZ^M\hS^{\mbox{\tiny e}}S^{\mbox{\tiny d}}}
	(u^K,x^M,y^M,z^M,s^{\mbox{\tiny e}},s^{\mbox{\tiny d}})=\bE\left\{P_{\hU^K\hX^M\hY^M\hZ^M\hS^{\mbox{\tiny e}}\hS^{\mbox{\tiny d}}}
	(u^K,x^M,y^M,z^M,s^{\mbox{\tiny e}},s^{\mbox{\tiny d}})\right\},
\end{equation}
where the expectation is w.r.t.\ both the randomness of the encoder and
the randomness of both channels. 
Note that

\begin{equation}
	P_{\hU^KX^MY^MZ^M\hS^{\mbox{\tiny e}}}
	(u^K,x^M,y^M,z^M,s^{\mbox{\tiny e}})=P_{\hU^K\hS^{\mbox{\tiny e}}}(u^K,s^{\mbox{\tiny e}})P(x^M|u^K,s^{\mbox{\tiny e}})
	Q_{\mbox{\tiny M}}(y^M|x^M)Q_{\mbox{\tiny W}}(z^M|y^M).
\end{equation}
where

\begin{eqnarray}
	P(x^M|u^K,s^{\mbox{\tiny e}})&=&\prod_{j=0}^{\ell-1} P(\tx_j|\tu_j,s_j^{\mbox{\tiny e}}),
	~~~~s_0^{\mbox{\tiny e}}=s^{\mbox{\tiny e}}\\
	Q_{\mbox{\tiny M}}(y^M|x^M)&=&\prod_{j=0}^{M-1}Q_{\mbox{\tiny M}}(y_i|x_i)\\
	Q_{\mbox{\tiny W}}(z^M|y^M)&=&\prod_{j=0}^{M-1}Q_{\mbox{\tiny M}}(z_i|y_i).
\end{eqnarray}
Note also that the bit error probability (in the absence of side information) under this distribution is

\begin{eqnarray}
	& &\frac{1}{K}\bE\{d_{\mbox{\tiny H}}(\hU^K,f(Y^M,S^{\mbox{\tiny d}}))\}\nonumber\\
	&=&\frac{1}{K}\sum_{u^K,y^M,s^{\mbox{\tiny e}},s^{\mbox{\tiny d}}}
	P_{\hU^KY^MS^{\mbox{\tiny e}}S^{\mbox{\tiny d}}}(u^K,y^M,s^{\mbox{\tiny e}},s^{\mbox{\tiny d}})
	d_{\mbox{\tiny H}}(u^K,f(y^M,s^{\mbox{\tiny d}}))\nonumber\\
	&=&\frac{1}{K}\sum_{u^K,y^M,s^{\mbox{\tiny d}}}\frac{K}{n}\sum_{i=0}^{n/K-1}
	\bE\bigg[\delta\{\tu_{i\ell}^{(i+1)\ell-1}=u^K,s_{i\ell+1}^{\mbox{\tiny e}}=s^{\mbox{\tiny e}},
	\ty_{i\ell}^{(i+1)\ell-1}=y^M,s_{i\ell+1}^{\mbox{\tiny d}}=s^{\mbox{\tiny d}}\bigg]\times\nonumber\\
	& &d_{\mbox{\tiny H}}(u^K,f(y^M,s^{\mbox{\tiny d}}))\nonumber\\
	&=&\frac{1}{n}\sum_{i=0}^{n/K-1}\sum_{y^M,s^{\mbox{\tiny d}}}F(y^M,s^{\mbox{\tiny d}}|
	u_{iK+1}^{iK+K},s_{i\ell+1}^{\mbox{\tiny e}})
	d_{\mbox{\tiny H}}(u_{iK+1}^{iK+K},f(y^M,s^{\mbox{\tiny d}}))\nonumber\\
	&=&\frac{1}{n}\sum_{i=1}^n\bE\{d_{\mbox{\tiny H}}(u_i,V_i)\},
\end{eqnarray}
where $f(Y^M,S^{\mbox{\tiny d}})$ is induced by $\ell$ successive applications of the decoder output function with inputs $Y^m, Y_{m+1}^{2m},\ldots,
Y_{M-m+1}^M$ and the initial state $S^{\mbox{\tiny d}}$, and where

\begin{equation}
	F(y^M,s^{\mbox{\tiny d}}|u^K,s^{\mbox{\tiny e}})=\sum_{x^M}P(x^M|u^K,s^{\mbox{\tiny e}})Q_{\mbox{\tiny M}}(y^M|x^M)
	P_{S^{\mbox{\tiny d}}|Y^M}(s^{\mbox{\tiny d}}|y^M).
\end{equation}

\subsection{Proof of Theorem \ref{thm1}}
\label{proof1}

Beginning with the reliability constraint, we have:

\begin{eqnarray}
        I(\hU^K;Y^M,S^{\mbox{\tiny d}})&=&H(\hU^K)-H(\hU^K|Y^M,S^{\mbox{\tiny d}})\nonumber\\
        &=&H(\hU^K)-H(\hU^K|Y^M)+I(S^{\mbox{\tiny d}};\hU^K|Y^M)\nonumber\\
        &\le&I(\hU^K;Y^M)+H(S^{\mbox{\tiny d}}|Y^M)\nonumber\\
        &\le&I(X^M;Y^M)+\log q_{\mbox{\tiny d}}.
\end{eqnarray}
On the other hand,

\begin{eqnarray}
	I(\hU^K;Y^M,S^{\mbox{\tiny d}})&=&H(\hU^K)-H(\hU^K|Y^M,S^{\mbox{\tiny d}})\nonumber\\
	&\ge&H(\hU^K)-K\Delta(\epsilon_{\mbox{\tiny r}}),
\end{eqnarray}
and so,

\begin{eqnarray}
	\label{rel1}
	I(X^M;Y^M)&\ge&H(\hU^K)-K\Delta(\epsilon_{\mbox{\tiny r}})-\log q_{\mbox{\tiny d}}\nonumber\\
	&\dfn&K\cdot R(u^n,q_{\mbox{\tiny d}},\epsilon_{\mbox{\tiny r}})\nonumber\\
	&=&M\cdot \frac{R(u^n,q_{\mbox{\tiny d}},\epsilon_{\mbox{\tiny r}})}{\lambda}.
\end{eqnarray}
Following \cite{Wyner75}, we define the function

\begin{equation}
	\Gamma[R]=\max_{\{P_X:~I(X;Y)\ge R\}}I(X;Y|Z)=
	\max_{\{P_X:~I(X;Y)\ge R\}}[I(X;Y)-I(X;Z)],
\end{equation}
which is monotonically non--increasing and concave \cite[Lemma 1]{Wyner75}.
Regarding the security constraint,

\begin{eqnarray}
        H(\hU^K)-K\epsilon_{\mbox{\tiny s}}&\lea&H(\hU^K)-\max_\mu I_\mu(U^K;Z^M)\nonumber\\
        &\le&H(\hU^K)-I(\hU^K;Z^M)\nonumber\\
	&=&H(\hU^K|Z^M)-H(\hU^K|Y^M,Z^M,S^{\mbox{\tiny d}})+H(\hU^K|Y^M,Z^M,S^{\mbox{\tiny d}})\nonumber\\
	&=&H(\hU^K|Z^M)-H(\hU^K|Y^M,Z^M)+I(S^{\mbox{\tiny d}};\hU^K|Y^M,Z^M)+H(\hU^K|Y^M,Z^M,S^{\mbox{\tiny d}})\nonumber\\
	&\leb&I(\hU^K;Y^M|Z^M)+\log q_{\mbox{\tiny d}}+K\Delta(\epsilon_{\mbox{\tiny r}})\nonumber\\
	&\lec&I(X^M;Y^M|Z^M)+\log q_{\mbox{\tiny d}}+K\Delta(\epsilon_{\mbox{\tiny r}})\nonumber\\
	&\led&\sum_{i=1}^MI(X_i;Y_i|Z_i,Y^{i-1})+\log q_{\mbox{\tiny d}}+K\Delta(\epsilon_{\mbox{\tiny r}})\nonumber\\
	&=&\sum_{i=1}^M\sum_{y^{i-1}}P_{Y^{i-1}}(y^{i-1})I(X_i;Y_i|Z_i,Y^{i-1}=y^{i-1})+
	\log q_{\mbox{\tiny d}}+K\Delta(\epsilon_{\mbox{\tiny r}})\nonumber\\
	&\lee&M\cdot\frac{1}{M}\sum_{i=1}^M\sum_{y^{i-1}}P_{Y^{i-1}}(y^{i-1})\Gamma[I(X_i;Y_i|Y^{i-1}=y^{i-1})]+
	\log q_{\mbox{\tiny d}}+K\Delta(\epsilon_{\mbox{\tiny r}})\nonumber\\
	&\lef&M\cdot\Gamma\bigg[\frac{1}{M}\sum_{i=1}^M\sum_{y^{i-1}}P_{Y^{i-1}}(y^{i-1})I(X_i;Y_i|Y^{i-1}=y^{i-1})\bigg]+
	\log q_{\mbox{\tiny d}}+K\Delta(\epsilon_{\mbox{\tiny r}})\nonumber\\
	&=&M\cdot\Gamma\bigg[\frac{1}{M}\sum_{i=1}^MI(X_i;Y_i|Y^{i-1})\bigg]+
	\log q_{\mbox{\tiny d}}+K\Delta(\epsilon_{\mbox{\tiny r}})\nonumber\\
	&=&M\cdot\Gamma\bigg[\frac{1}{M}\sum_{i=1}^M\bigg\{H(Y_i|Y^{i-1})-H(Y_i|X_i,Y^{i-1})\bigg\}\bigg]+
	\log q_{\mbox{\tiny d}}+K\Delta(\epsilon_{\mbox{\tiny r}})\nonumber\\
	&=&M\cdot\Gamma\bigg[\frac{1}{M}\sum_{i=1}^M\bigg\{H(Y_i|Y^{i-1})-H(Y_i|X_i)\bigg\}\bigg]+
	\log q_{\mbox{\tiny d}}+K\Delta(\epsilon_{\mbox{\tiny r}})\nonumber\\
	&=&M\cdot\Gamma\bigg[\frac{1}{M}\bigg\{H(Y^M)-H(Y^M|X^M)\bigg\}\bigg]+
	\log q_{\mbox{\tiny d}}+K\Delta(\epsilon_{\mbox{\tiny r}})\nonumber\\
	&=&M\cdot\Gamma\bigg[\frac{I(X^M;Y^M)}{M}\bigg]+
	\log q_{\mbox{\tiny d}}+K\Delta(\epsilon_{\mbox{\tiny r}})\nonumber\\
	&\leg&M\cdot\Gamma\bigg[\frac{R(u^n,q_{\mbox{\tiny d}},\epsilon_{\mbox{\tiny r}})}{\lambda}\bigg]+
	\log q_{\mbox{\tiny d}}+K\Delta(\epsilon_{\mbox{\tiny r}})\nonumber\\
	&\le&M\cdot\Gamma\bigg[\frac{R(u^n,q_{\mbox{\tiny d}},\epsilon_{\mbox{\tiny r}})-\epsilon_{\mbox{\tiny s}}}{\lambda}\bigg]+
	\log q_{\mbox{\tiny d}}+K\Delta(\epsilon_{\mbox{\tiny r}}),
\end{eqnarray}
where $P_{Y^{i-1}}(y^{i-1})=\sum_{y_i^M}P_{Y^M}(y^M)$,
(a) is due to the security constraint, (b) follows from Fano's inequality and the fact that
$I(S^{\mbox{\tiny d}};\hU^K|Y^M,Z^M)\le H(S^{\mbox{\tiny d}})\le \log q_{\mbox{\tiny d}}$, 
(c) is by the data processing inequality and the fact that $\hU^K\to X^M\to Y^M$ is a Markov chain given $Z^M$,
(d) is as in \cite[eq.\ (37)]{Wyner75}, (e) is by the definition of Wyner's function $\Gamma(\cdot)$, (f) is by the concavity of this
function, and (g) is by (\ref{rel1}) and the decreasing monotonicity of the function $\Gamma(\cdot)$.
Thus,

\begin{equation}
	\frac{R(u^n,q_{\mbox{\tiny d}},\epsilon_{\mbox{\tiny r}})-\epsilon_{\mbox{\tiny s}}}{\lambda}\le
	\Gamma\bigg[\frac{R(u^n,q_{\mbox{\tiny d}},\epsilon_{\mbox{\tiny r}})-\epsilon_{\mbox{\tiny s}}}{\lambda}\bigg]
\end{equation}
or

\begin{equation}
	\frac{R(u^n,q_{\mbox{\tiny d}},\epsilon_{\mbox{\tiny r}})-\epsilon_{\mbox{\tiny s}}}{\lambda}\le C_{\mbox{\tiny s}}
\end{equation}
which is

\begin{equation}
	R(u^n,q_{\mbox{\tiny d}},\epsilon_{\mbox{\tiny r}})\le \lambda C_{\mbox{\tiny s}}+\epsilon_{\mbox{\tiny s}}
\end{equation}
or, equivalently,

\begin{equation}
	\frac{H(\hU^K)}{K}\le \lambda C_{\mbox{\tiny s}}+\epsilon_{\mbox{\tiny s}}+\Delta(\epsilon_{\mbox{\tiny r}})+
	\frac{\log q_{\mbox{\tiny d}}}{K}.
\end{equation}
Finally, we apply the inequality \cite[eq.\ (18)]{me13},

\begin{equation}
	\frac{H(\hU^K)}{K}\ge \rho_{\mbox{\tiny LZ}}(u^n)-\frac{2K(\log\alpha+1)^2}{(1-\epsilon_n)\log n}-\frac{2K\alpha^{2K}\log\alpha}{n}-\frac{1}{K},
\end{equation}
to obtain

\begin{equation}
	\rho_{\mbox{\tiny LZ}}(u^n)\le 
	\lambda C_{\mbox{\tiny s}}+\epsilon_{\mbox{\tiny s}}+\Delta(\epsilon_{\mbox{\tiny r}})+\zeta_n(q_{\mbox{\tiny d}},k),
\end{equation}
which completes the proof of Theorem \ref{thm1}.

\subsection{Proof of Theorem \ref{thm2}}

Consider the following extension of the joint distribution to include a random variable that represents $\{b_i\}$, as
follows:

\begin{eqnarray}
        & &P_{\hU^KB^JX^MY^MZ^M\hS^{\mbox{\tiny e}}S^{\mbox{\tiny d}}}
        (u^K,b^J,x^M,y^M,z^M,s^{\mbox{\tiny e}},s^{\mbox{\tiny d}})=\nonumber\\
	& &\frac{K}{n}\sum_{i=0}^{n/K-1}\bE\bigg[\delta\{\tu_{i\ell}^{(i+1)\ell-1}=u^K,\tb_{i\ell}^{(i+1)\ell-1}=b^J,
        \tx_{i\ell}^{(i+1)\ell-1}=x^M,\ty_{i\ell}^{(i+1)\ell-1}=y^M,\nonumber\\
        & &\tz_{i\ell}^{(i+1)\ell-1}=z^M,s_{i\ell+1}^{\mbox{\tiny e}}=s^{\mbox{\tiny e}},
	s_{i\ell+1}^{\mbox{\tiny d}}=s^{\mbox{\tiny d}}\}\bigg],
\end{eqnarray}
where $J=j\ell$ and $\tb_{i\ell}^{(i+1)\ell-1}=(\tb_{i\ell},\tb_{i\ell+1},\ldots,\tb_{(i+1)\ell-1})$.
Next, consider the following chain of inequalities

\begin{eqnarray}
	K\epsilon_{\mbox{\tiny s}}&\ge&\max_\mu I_\mu(U^K;Z^M)\nonumber\\
	&\ge&I(\hU^K;Z^M)\nonumber\\
	&=&I(\hU^K,B^J,S^{\mbox{\tiny e}};Z^M)-I(B^J,S^{\mbox{\tiny e}};Z^M|\hU^K)\nonumber\\
	&\eqa&I(X^M;Z^M)-I(B^J,S^{\mbox{\tiny e}};Z^M|\hU^K)\nonumber\\
	&\ge&I(X^M;Z^M)-H(B^J,S^{\mbox{\tiny e}}|\hU^K)\nonumber\\
	&\ge&I(X^M;Z^M)-H(B^J,S^{\mbox{\tiny e}})\nonumber\\
	&\ge&I(X^M;Z^M)-H(B^J)-H(S^{\mbox{\tiny e}})\nonumber\\
	&\ge&I(X^M;Z^M)-J-\log q_{\mbox{\tiny e}},
\end{eqnarray}
where (a) is due to the fact that, on the one hand, $X^M$ is a deterministic function of $(\hU^K,B^J,S^{\mbox{\tiny e}})$,
which implies that $I(\hU^K,B^J,S^{\mbox{\tiny e}};Z^M)\ge I(X^M;Z^M)$, but on the other hand,
$(\hU^K,B^J,S^{\mbox{\tiny e}})\to X^M\to Z^M$ is a Markov chain and so,
$I(\hU^K,B^J,S^{\mbox{\tiny e}};Z^M)\le I(X^M;Z^M)$, hence the equality.
Thus,

\begin{equation}
	J\ge I(X^M;Z^M)- K\epsilon_{\mbox{\tiny s}}-\log q_{\mbox{\tiny e}},
\end{equation}
or

\begin{equation}
	j\ge \frac{I(X^M;Z^M)}{\ell}- k\epsilon_{\mbox{\tiny s}}-\frac{\log q_{\mbox{\tiny e}}}{\ell}
	\ge \frac{mI(X^M;Z^M)}{M}- k\epsilon_{\mbox{\tiny s}}-\frac{\log q_{\mbox{\tiny e}}}{\ell}.
\end{equation}
The meaning of this result is the following: once one finds a communication system that complies with both the
security constraint and the reliability constraint, then the amount of local randomization is lower bounded in terms of the
induced mutual information, $I(X^M;Z^M)$, as above. By the hypothesis of Theorem \ref{thm2}, the secrecy capacity is saturated,
and hence $P_{X^M}$ must coincide with the product distribution, $[P_{X^*}]^M$, yielding $I(X^M;Z^M)/M=I(X^*;Z^*)$. Thus,

\begin{equation}
	j\ge mI(X^*;Z^*)- k\epsilon_{\mbox{\tiny s}}-\frac{\log q_{\mbox{\tiny e}}}{\ell}.
\end{equation}
This completes the proof of Theorem \ref{thm2}.

\subsection*{Outline of the Proof of Theorem \ref{thm3}}

The proof follows essentially 
the same steps as those of the proof of Theorem \ref{thm1}, except that everything should be conditioned on the side information,
but there are also some small twists. We will therefore only provide a proof outline and highlight the differences.

The auxiliary joint distribution is now extended to read

\begin{eqnarray}
& &P_{\hU^K\hW^K\dW^KX^NY^NZ^N\hS^{\mbox{\tiny e}}S^{\mbox{\tiny d}}}
        (u^K,w^K,\dw^K,x^N,y^N,z^N,s^{\mbox{\tiny e}},s^{\mbox{\tiny d}})=\nonumber\\
	& &\frac{K}{m}\sum_{i=0}^{m/K-1}\bE\bigg[\delta\{\tu_{i\ell}^{(i+1)\ell-1}=u^K,\tw_{i\ell}^{(i+1)\ell-1}=w^K,
	\tilde{\dw}_{i\ell}^{(i+1)\ell-1}=\dw^K,\tx_{i\ell}^{(i+1)\ell-1}=x^M,\nonumber\\
	& &\ty_{i\ell}^{(i+1)\ell-1}=y^M,
        \tz_{i\ell}^{(i+1)\ell-1}=z^M,s_{i\ell+1}^{\mbox{\tiny e}}=s^{\mbox{\tiny e}},
	s_{i\ell+1}^{\mbox{\tiny d}}=s^{\mbox{\tiny d}}\}\bigg].
\end{eqnarray}
Note that

\begin{eqnarray}
	& &P_{\hU^K\hW^K\dW^KZ^M\hS^{\mbox{\tiny e}}}(u^k,w^k,\dw^k,z^M,s^{\mbox{\tiny e}})\nonumber\\
	&=&\frac{K}{n}\sum_{i=0}^{n/K-1}\bE\bigg[\delta\{\tu_{i\ell}^{(i+1)\ell-1}=u^K,\tw_{i\ell}^{(i+1)\ell-1}=w^K,
	 \tilde{\dw}_{i\ell}^{(i+1)\ell-1}=\dw^K,\tz_{i\ell}^{(i+1)\ell-1}=z^M,
	 s_{i\ell+1}^{\mbox{\tiny e}}=s^{\mbox{\tiny e}}\}\bigg]\nonumber\\
	&=&\frac{K}{n}\sum_{i=0}^{n/K-1}\delta\{\tu_{i\ell}^{(i+1)\ell-1}=u^K,
	\tw_{i\ell}^{(i+1)\ell-1}=w^K,s_{i\ell+1}^{\mbox{\tiny e}}=s^{\mbox{\tiny e}}\}\cdot Q_{\dW^K|W^K}(\dw^K|w^K)\cdot
	G(z^M|u^K,s^{\mbox{\tiny e}})\nonumber\\
	&=&P_{\hU^K\hW^K\hS^{\mbox{\tiny e}}}(u^K,w^K,s^{\mbox{\tiny e}})\cdot Q_{\dW^K|W^K}(\dw^K|w^K)\cdot 
	G(z^M|u^K,s^{\mbox{\tiny e}}),
\end{eqnarray}
where

\begin{equation}
	G(z^M|u^K,s^{\mbox{\tiny e}})=\sum_{x^M}P(x^M|u^K,s^{\mbox{\tiny e}})Q_{\mbox{\tiny MW}}(z^M|x^M).
\end{equation}
It follows that $\dW^K\to \hW^K\to (\hU^K,\hS^{\mbox{\tiny e}})\to Z^M$ is a Markov chain under
$P_{\hU^K\hW^K\dW^KZ^M\hS^{\mbox{\tiny e}}}$. In other words, the legitimate decoder has side information of better quality than 
that of the wiretapper.
First, observe that

\begin{eqnarray}
	\label{switchsi}
	I_\mu(U^n;Z^N|W^n)&=&H_\mu(Z^N|W^n)-H_\mu(Z^N|W^n,U^n)\nonumber\\
	&\le&H_\mu(Z^N|W^n)-H_\mu(Z^N|W^n,U^n,S^{\mbox{\tiny e}})\nonumber\\
	&=&H_\mu(Z^N|W^n)-H_\mu(Z^N|U^n,S^{\mbox{\tiny e}})\nonumber\\
	&\le&H_\mu(Z^N|\dW^n)-H_\mu(Z^N|U^n,S^{\mbox{\tiny e}})\nonumber\\
	&=&H_\mu(Z^N|\dW^n)-H_\mu(Z^N|\dW^n,U^n,S^{\mbox{\tiny e}})\nonumber\\
	&\le&H_\mu(Z^N|\dW^n)-H_\mu(Z^N|\dW^n,U^n)+\log q_{\mbox{\tiny e}}\nonumber\\
	&=&I_\mu(U^n;Z^N|\dW^n)+\log q_{\mbox{\tiny e}}.
\end{eqnarray}
The reliability constraint is handled exactly as in the proof of Theorem \ref{thm1}, except
that everything should be conditioned on $\hW^K$. The result of this is

\begin{eqnarray}
        I(X^M;Y^M|\hW^K)&\ge&H(\hU^K|\hW^K)-K\Delta(\epsilon_{\mbox{\tiny r}})-\log q_{\mbox{\tiny d}}\nonumber\\
        &\dfn&K\cdot R(u^n,w^n,q_{\mbox{\tiny d}},\epsilon_{\mbox{\tiny r}})\nonumber\\
        &=&M\cdot \frac{R(u^n,w^n,q_{\mbox{\tiny d}},\epsilon_{\mbox{\tiny r}})}{\lambda}.
\end{eqnarray}

Regarding the security constraint, we begin with the following manipulation.

\begin{eqnarray}
H(\hU^K|Z^M,\dW^K)&=&H(\hU^K|\dW^K)-I(\hU^K;Z^M|\dW^K)\nonumber\\
&=&H(\hU^K|\dW^K)-H(\hU^K|\hW^K)+H(\hU^K|\hW^K)-I(\hU^K;Z^M|\dW^K)\nonumber\\
&\lea&H(\hU^K|\dW^K)-H(\hU^K|\hW^K)+H(\hU^K|\hW^K)-I(\hU^K;Z^M|\hW^K)+\log q_{\mbox{\tiny e}}\nonumber\\
&=&H(\hU^K|\dW^K)-H(\hU^K|\hW^K)+H(\hU^K|Z^M,\hW^K)+\log q_{\mbox{\tiny e}}\nonumber\\
&=&H(\hU^K|\dW^K)-H(\hU^K|\hW^K)+
H(\hU^K|Z^M,\hW^K)-\nonumber\\
& &H(\hU^K|Y^M,Z^M,S^{\mbox{\tiny d}},\hW^K)+
H(\hU^K|Y^M,Z^M,S^{\mbox{\tiny d}},\hW^K)+\log q_{\mbox{\tiny e}}\nonumber\\
&=&H(\hU^K|\dW^K)-H(\hU^K|\hW^K)+H(\hU^K|Z^M,\hW^K)-H(\hU^K|Y^M,Z^M,\hW^K)+\nonumber\\
& &I(S^{\mbox{\tiny d}};\hU^K|Y^M,Z^M,\hW^K)+
H(\hU^K|Y^M,Z^M,S^{\mbox{\tiny d}},\hW^K)+\log q_{\mbox{\tiny e}}\nonumber\\
&\leb&H(\hU^K|\dW^K)-H(\hU^K|\hW^K)+I(\hU^K;Y^M|Z^M,\hW^K)+\log q_{\mbox{\tiny d}}+\nonumber\\
& &K\Delta(\epsilon_{\mbox{\tiny r}})+\log q_{\mbox{\tiny e}}\nonumber\\
&\le&H(\hU^K|\dW^K)-H(\hU^K|\hW^K)+I(\hU^K,\hS^{\mbox{\tiny e}};Y^M|Z^M,\hW^K)+\log q_{\mbox{\tiny d}}+\nonumber\\
& &K\Delta(\epsilon_{\mbox{\tiny r}})+\log q_{\mbox{\tiny e}}\nonumber\\
&\lec&H(\hU^K|\dW^K)-H(\hU^K|\hW^K)+I(X^M;Y^M|Z^M,\hW^K)+\log(q_{\mbox{\tiny e}}q_{\mbox{\tiny d}})+
K\Delta(\epsilon_{\mbox{\tiny r}}),
\end{eqnarray}
where in (a) we have used eq.\ (\ref{switchsi}), 
in (b) we used Fano's inequality, and in (c) we used the data processing inequality as $(\hU^K,\hS^{\mbox{\tiny e}})\to X^M\to Y^M$ is a Markov
chain (also conditioned on $(\hW^K,Z^M)$). The next step is to further upper bound the term $I(X^M;Y^M|Z^M,\hW^K)$. This is carried out
very similarly as in the proof of Theorem \ref{thm1}, except that everything is conditioned also on $\hW^K$. We then obtain

\begin{eqnarray}
H(\hU^K|Z^M,\dW^K)&\le&H(\hU^K|\dW^K)-H(\hU^K|\hW^K)+
M\cdot\Gamma\bigg[\frac{R(u^n,w^n,q_{\mbox{\tiny d}},\epsilon_{\mbox{\tiny r}})}{\lambda}\bigg]+\nonumber\\
& &\log(q_{\mbox{\tiny e}}q_{\mbox{\tiny d}})+K\Delta(\epsilon_{\mbox{\tiny r}}),
\end{eqnarray}
or, equivalently,

\begin{eqnarray}
& &H(\hU^K|\hW^K)-M\cdot\Gamma\bigg[\frac{R(u^n,w^n,q_{\mbox{\tiny d}},\epsilon_{\mbox{\tiny r}})}{\lambda}\bigg]\nonumber\\
&\le&H(\hU^K|\dW^K)-H(\hU^K|Z^M,\dW^K)+\log(q_{\mbox{\tiny e}}q_{\mbox{\tiny d}})+K\Delta(\epsilon_{\mbox{\tiny r}})\nonumber\\
&=&I(\hU^K;Z^M|\dW^K)+\log(q_{\mbox{\tiny e}}q_{\mbox{\tiny d}})+K\Delta(\epsilon_{\mbox{\tiny r}})\nonumber\\
&\le&K\epsilon_{\mbox{\tiny s}}+\log(q_{\mbox{\tiny e}}q_{\mbox{\tiny d}})+K\Delta(\epsilon_{\mbox{\tiny r}}),
\end{eqnarray}
or

\begin{eqnarray}
R(u^n,w^n,q_{\mbox{\tiny e}}\cdot q_{\mbox{\tiny d}},\epsilon_{\mbox{\tiny r}})&\le&\lambda\cdot\Gamma\bigg[
\frac{R(u^n,w^n,q_{\mbox{\tiny d}},\epsilon_{\mbox{\tiny r}})}{\lambda}\bigg]\nonumber\\
&\le&\lambda\cdot\Gamma\bigg[
\frac{R(u^n,w^n,q_{\mbox{\tiny e}}\cdot q_{\mbox{\tiny d}},\epsilon_{\mbox{\tiny r}})}{\lambda}\bigg]\nonumber\\
\end{eqnarray}
which is the same as

\begin{equation}
R(u^n,w^n,q_{\mbox{\tiny e}}\cdot q_{\mbox{\tiny d}},\epsilon_{\mbox{\tiny r}})\le\lambda\cdot C_{\mbox{\tiny s}}.
\end{equation}
or

\begin{equation}
\frac{H(\hU^K|\hW^K)}{K}\le\lambda\cdot C_{\mbox{\tiny s}}+\epsilon_{\mbox{\tiny s}}+\Delta(\epsilon_{\mbox{\tiny r}})+
\frac{\log(q_{\mbox{\tiny e}}\cdot q_{\mbox{\tiny d}})}{K}.
\end{equation}
The proof is completed by combining the last inequality with the following inequality \cite[eqs.\ (17)-(19)]{me00}, \cite[eqs.\ (55)--57)]{me20}:

\begin{equation}
	\frac{H(\hU^K|\hW^K)}{K}\ge\rho_{\mbox{\tiny LZ}}(u^n|w^n)-\frac{\log(4A^2)}{(1-\epsilon_n)\log n}-\frac{A^2\log(4A^2)}{n}-\frac{1}{K},
\end{equation}
where $A=[(\alpha\omega)^{K+1}-1]/[\alpha\omega-1]$, $\omega$ being the alphabet size of $\calW$.\\

\section*{Acknowledgment}
Interesting discussions with Alejandro Cohen are acknowledged with thanks.

\end{document}